\documentclass[sn-aps]{sn-jnl}
\jyear{2022}
\usepackage{caption}
\usepackage{subcaption}
\usepackage{adjustbox}
\usepackage{verbatim}
\usepackage[left]{lineno}
\usepackage{makecell}

\begin{document}
\title[ ]{Measurements of efficiency, timing  and response to irradiation for direct detection of charged particles with SiPMs}

\author*[1,6]{\fnm{F.} \sur{Carnesecchi}}\email{francesca.carnesecchi@cern.ch}
\author*[2,3]{\fnm{B.} \sur{Sabiu}}\email{b.sabiu@unibo.it}
\author[2,3]{\fnm{B.R.} \sur{Achari}}
\author[2,3]{\fnm{N.} \sur{Agrawal}}
\author[2,3]{\fnm{A.} \sur{Alici}}
\author[3]{\fnm{P.} \sur{Antonioli}}
\author[2,3]{\fnm{S.} \sur{Arcelli}}
\author[3]{\fnm{C.} \sur{Baldanza}}
\author[2,3]{\fnm{F.} \sur{Bellini}}
\author[4,7]{\fnm{S.} \sur{Bufalino}}
\author[3]{\fnm{D.} \sur{Cavazza}}
\author[2,3]{\fnm{L.} \sur{Cifarelli}}
\author[9,3]{\fnm{G.} \sur{Clai}}
\author[2,3]{\fnm{M.} \sur{Colocci}}
\author[4,7]{\fnm{S.} \sur{Durando}}
\author[2,3]{\fnm{F.} \sur{Ercolessi}}
\author[2,3]{\fnm{G.} \sur{Fabbri}}
\author[3]{\fnm{D.} \sur{Falchieri}}
\author[7]{\fnm{C.} \sur{Ferrero}}
\author[5]{\fnm{A.} \sur{Ficorella}}
\author[4,7]{\fnm{U.} \sur{Follo}}
\author[1,3]{\fnm{M.} \sur{Garbini}}
\author[2,3]{\fnm{S.} \sur{Geminiani}}
\author[10]{\fnm{G.} \sur{Gioachin}}
\author[5]{\fnm{A.} \sur{Gola}}
\author[3]{\fnm{D.} \sur{Hatzifotiadou}}
\author[3]{\fnm{A.} \sur{Khuntia}}
\author[3]{\fnm{I.} \sur{Lax}}
\author[2]{\fnm{M.} \sur{Maestrelli}}
\author[3]{\fnm{A.} \sur{Margotti}}
\author[2,3]{\fnm{G.} \sur{Malfattore}}
\author[3]{\fnm{R.} \sur{Nania}}
\author[3]{\fnm{F.} \sur{Noferini}}
\author[5]{\fnm{L.} \sur{Parellada-Monreal}}
\author[5]{\fnm{M.} \sur{Penna}}
\author[3]{\fnm{O.} \sur{Pinazza}}
\author[3]{\fnm{R.} \sur{Preghenella}}
\author[2,3]{\fnm{M.} \sur{Razza}}
\author[3]{\fnm{R.} \sur{Ricci}}
\author[3]{\fnm{L.} \sur{Rignanese}}
\author[4,7]{\fnm{A.} \sur{Rivetti}}
\author[2,3]{\fnm{G.} \sur{Romanenko}}
\author[2,3]{\fnm{N.} \sur{Rubini}}
\author[2,3]{\fnm{E.} \sur{Rovati}}
\author[3]{\fnm{E.} \sur{Scapparone}}
\author[2,3]{\fnm{G.} \sur{Scioli}}
\author[2,3]{\fnm{S.} \sur{Strazzi}}
\author[2,3]{\fnm{S.} \sur{Tomassini}}
\author[8]{\fnm{C.} \sur{Veri}}
\author[2,3]{\fnm{A.} \sur{Zichichi}}

\affil[1]{ \orgname{Museo Storico della Fisica e Centro Studi Enrico Fermi}, \orgaddress{\street{Via Panisperna 89 A}, \city{Roma}, \postcode{10129}, \country{Italy}}}

\affil[2]{\orgdiv{Dipartimento di Fisica e Astronomia "A. Righi"}, \orgname{University of Bologna}, \orgaddress{\street{viale Carlo Berti Pichat 6/2}, \city{Bologna}, \postcode{40127}, \country{Italy}}}

\affil[3]{\orgdiv{Sezione di Bologna}, \orgname{Istituto Nazionale di Fisica Nucleare}, \orgaddress{\street{viale Carlo Berti Pichat 6/2}, \city{Bologna}, \postcode{40127}, \country{Italy}}}

\affil[4]{\orgdiv{Dipartimento di Elettronica e Telecomunicazioni (DET)}, \orgname{Politecnico di
Torino}, \orgaddress{\street{Corso Duca degli Abruzzi, 24}, \city{Torino}, \postcode{10129}, \country{Italy}}}

\affil[5]{\orgname{Fondazione Bruno Kessler}, \orgaddress{\street{Via Sommarive, 18}}, \orgaddress{\city{Povo}, \postcode{38123}, \country{Italy}}}


\affil[6]{\orgdiv{Laboratori Nazionali di Frascati}, \orgname{Istituto Nazionale di Fisica Nucleare}, \orgaddress{\street{Via Enrico Fermi 54}, \city{Roma}, \postcode{00044}, \country{Italy}}}

\affil[7]{\orgdiv{Sezione di Torino}, \orgname{Istituto Nazionale di Fisica Nucleare}, \orgaddress{\street{Via Pietro Giuria, 1}, \city{Torino}, \postcode{10125}, \country{Italy}}}

\affil[8]{\orgdiv{Sezione di Lecce}, \orgname{Istituto Nazionale di Fisica Nucleare}, \orgaddress{\street{Via Provinciale per Arnesano}, \city{Lecce}, \postcode{73100}, \country{Italy}}}

\affil[9]{\orgdiv{Sede di Bologna}, \orgname{Italian National Agency for New Technologies, Energy and Sustainable Economic Development (ENEA)}, \orgaddress{\street{Via dei Mille 21}, \city{Bologna}, \postcode{40121}, \country{Italy}}}
\affil[10]{\orgdiv{Dipartimento di Scienza Applicata e Tecnologia (DISAT)}, \orgname{Politecnico di
Torino}, \orgaddress{\street{Corso Duca degli Abruzzi, 24}, \city{Torino}, \postcode{10129}, \country{Italy}}}


\abstract{In this paper the efficiency of direct charged particle detection with different Silicon PhotoMultiplier (SiPM) sensors has been measured to be close to 100\%.
Time resolution of about 20 ps has also been confirmed for sensors with an active area of around 3$\times$3 mm$^\text{2}$ and a single-cell area of 40 \textmu m$^\text{2}$.
In addition, the SiPM performance after irradiation, in terms of timing response and dark count rate, has been evaluated for sensors with a 1$\times$1 mm$^\text{2}$ area, demonstrating that SiPMs can maintain excellent timing capabilities and a low dark count rate when an appropriate threshold is applied to the signal.
}


\maketitle

\section{Introduction}\label{sec1}

Silicon PhotoMultipliers (SiPMs) are conventionally coupled to external scintillators or Cherenkov radiators. 
However, in  \cite{SiPM1, SiPM2, SiPM3} a different approach was quantitatively demonstrated: direct detection of charged particles using only the Cherenkov radiation produced within the standard, factory-supplied protection layer of the SiPM itself.
This simple choice yields a very high number of firing cells (SPADs) leading to an excellent time resolution of around 20 ps for events with more than 5 SPADs firing, which constituted more than $\sim$90\% of the cases at 4 V of OverVoltage ($V_{\text{OV}}$).

The large number of SPADs firing indicated potentially high efficiency, which was preliminarily studied in \cite{SiPM1}. In the present paper, a more detailed study on the efficiency has been performed thanks to the availability of SiPMs with a larger active area with respect to \cite{SiPM2,SiPM3}, that allowed to cover all the surface subtended by the Cherenkov light cone. 

In addition to this study, sensors used in \cite{SiPM3} were tested after an irradiation campaign at the INFN TIFPA\footnote{\href{https://www.tifpa.infn.it/}{https://www.tifpa.infn.it/}} facility in Trento (Italy). The measurements were done in terms of Dark Count Rate (DCR) and timing performances with respect to an increasing number of firing SPADs.

This work is part of the R\&D studies for the Time Of Flight (TOF) layers of the ALICE 3 \cite{ALICE3} experiment at the LHC.

\section{Experimental setup}\label{sec2}

\subsection{Detectors} 
\label{sec:detectors}
For the present study available NUV-HD-LFv2 and NUV-HD-RH SiPMs produced by Fondazione Bruno Kessler (FBK) were used  \cite{2020Mazzi, Altamura, ALTAMURA2023, 2019gola}.

\begin{table}[h]
\caption{Main features of the SiPMs under test. The thickness of the protection layer is measured from the board hosting the sensor i.e. the effective thickness is given by the protection layer minus 550 \textmu m (sensor). SR = Silicone Resin, WR = Without Resin, n = refraction index.
}
\label{tab:names}
\centering
\setlength{\tabcolsep}{3pt}
\begin {tabular}{ l cc cc ccc }
\toprule
\multirow{2}{*}[-0.8ex]{ } & \multirow{2}{*}[-0.8ex]{\textbf{Area (mm$^2$)}} & \multirow{2}{*}[-0.8ex]{\textbf{FF (\%)}} & \multicolumn{2}{c}{\textbf{SPAD}} & \multicolumn{3}{c}{\textbf{Protection layer}} \\
\cmidrule(lr){4-5} \cmidrule(lr){6-8}
\cmidrule(lr){4-5} \cmidrule(lr){6-8}
& & & \textbf{Pitch} & \textbf{\#} & \textbf{Type} & \textbf{Thick} & \textbf{n} \\
\midrule
\textbf{SR3-3x3-40} & & & & & Silicone & 3 mm & 1.52 \\
\textbf{SR15-3x3-40} & 3.20$\times$3.12 & 83 & 40 \textmu m & 6200 & Silicone & 1.5 mm & 1.52 \\
\textbf{WR-3x3-40} & & & & & \multicolumn{3}{c}{Without resin} \\
\midrule
\textbf{SR1-1x1-20} & 1$\times$1 & 72 & 20 \textmu m & 2444 & Silicone & 1 mm & 1.50 \\
\bottomrule
\end{tabular}
\end{table}

Two types of SiPMs, whose names and main features are reported in Table~\ref{tab:names}, were under study:
\begin{itemize}
    \item \textbf{3x3-40}: used in the efficiency and timing measurements. They feature NUV-HD-LFv2 technology, square SPADs and breakdown voltage V$_{bd}$ = 32.2 $\pm$0.1 V \cite{2019gola, Gundacker_2023}.
    \item \textbf{1x1-20}: same used in \cite{SiPM3} but compared in terms of DCR and timing response before and after a proton irradiation at TIFPA of 10$^{10}$~1~MeV~n$_\text{eq}$~cm$^{-2}$. These SiPMs feature NUV-HD-RH technology, hexagonal SPADs and V$_{bd}$ = 33.0 $\pm$ 0.1 V.
\end{itemize}

The SiPMs were produced both with and without a silicone resin protection layer of different thicknesses.
Since the sensors are 550 \textmu m thick, protection layers of 1, 1.5, and 3 mm correspond to resin thicknesses of 450, 950, and 2450 \textmu m above the sensor surface, respectively.


\subsection{Beam test setup} 
\label{subsec:tb}
The SiPMs were tested at the CERN–PS T10,  with a particle beam of 10~GeV/$c$  mainly composed of protons and $\pi^+$.

The telescope was made of four sensors: two SiPMs under test and two  LGAD detectors (1x1 mm$^2$ area and 35~\textmu  m or 25~\textmu m thick sensors) \cite{LGAD}. The latter were used as timing reference and positioned one at the beginning and one at the end of the telescope with respect to the beam direction. The LGADs define a trigger for the beam particles given by their coincidence. The four sensors were placed at a distance of around 8 cm from each other. Each detector was mounted on a remotely controlled moving stage capable of a 10~\textmu m precision positioning in both directions orthogonal to the beam axis: this configuration allowed  a precise alignment of the SiPM with respect to the LGAD trigger. 

The whole setup was enclosed in a dark box. The sensors were operated with a simple Peltier cell with water cooling and the temperature ranged between 20°C and 25°C. 

The SiPM signals were independently coupled to a customized front-end with X-LEE amplifiers\footnote{\href{https://www.minicircuits.com/pdfs/LEE-39+.pdf}{https://www.minicircuits.com/pdfs/LEE-39+.pdf}}  featuring a total gain factor of about 40 dB. At each trigger, all four waveforms were stored using a Lecroy Wave-Runner 9404M-MS digital oscilloscope.

%


\subsection{Analysis method}
\label{subsec:signal}

Signal and noise events selection from raw waveforms proceeds through different steps, similar to the procedure described in \cite{SiPM3}.\\
The $\pm$2 ns window centered on the trigger time (given by the coincidence of the two LGAD sensors) defines the temporal location of the expected SiPM signal.\\
The baseline of every signal is evaluated on an event-by-event basis over a time window of 3 ns 
immediately preceding the signal window and is subtracted from the entire waveform.
The DCR is computed on the same interval.\\
A selection was implemented to reject events where the tail of a preceding signal could contaminate the baseline and signal amplitude measurements: we rejected events where the amplitude within a 1 ns time window located 4 ns prior to the signal window exceeded $3\sigma$ of the baseline.
This cut was necessary (unlike in previous studies \cite{SiPM1, SiPM2, SiPM3}) due to the high particle beam rate ($\sim$ tens of kHz/mm$^2$) and the large-area SiPMs, particularly when combined with amplifier saturation.


The efficiency has been studied for the 3x3-40 SiPM\footnote{Notice that the trigger area defined by the LGADs is much smaller than the sensitive area of the SiPM.} as a function of a variable threshold on the signal rising edge.
This analysis is restricted to the selected events and the efficiency is defined as:
\begin{equation}
\text{Efficiency} = \frac{\text{number of events with SiPM signal above threshold}}{\text{total number of events}}
\end{equation}
The noise was determined analogously but within the baseline evaluation window. This allows us to quantify the false positive rate (or fake hit rate) related to the threshold setting.

For the timing analysis, the measurements were always performed using the LGAD as a reference. For the latter, a CFD (Constant Fraction Discrimination) threshold of 50$\%$ is used. The reference LGAD time resolution was  measured in the same test beam to be $\sim 26$  ps \cite{LGAD}. 



\section{Results}
\label{result}

\subsection{Efficiency and timing measurements}
\label{subsec:efficiency}
\begin{figure}[h!]
        \centering%
        \includegraphics [width=0.6\textwidth]{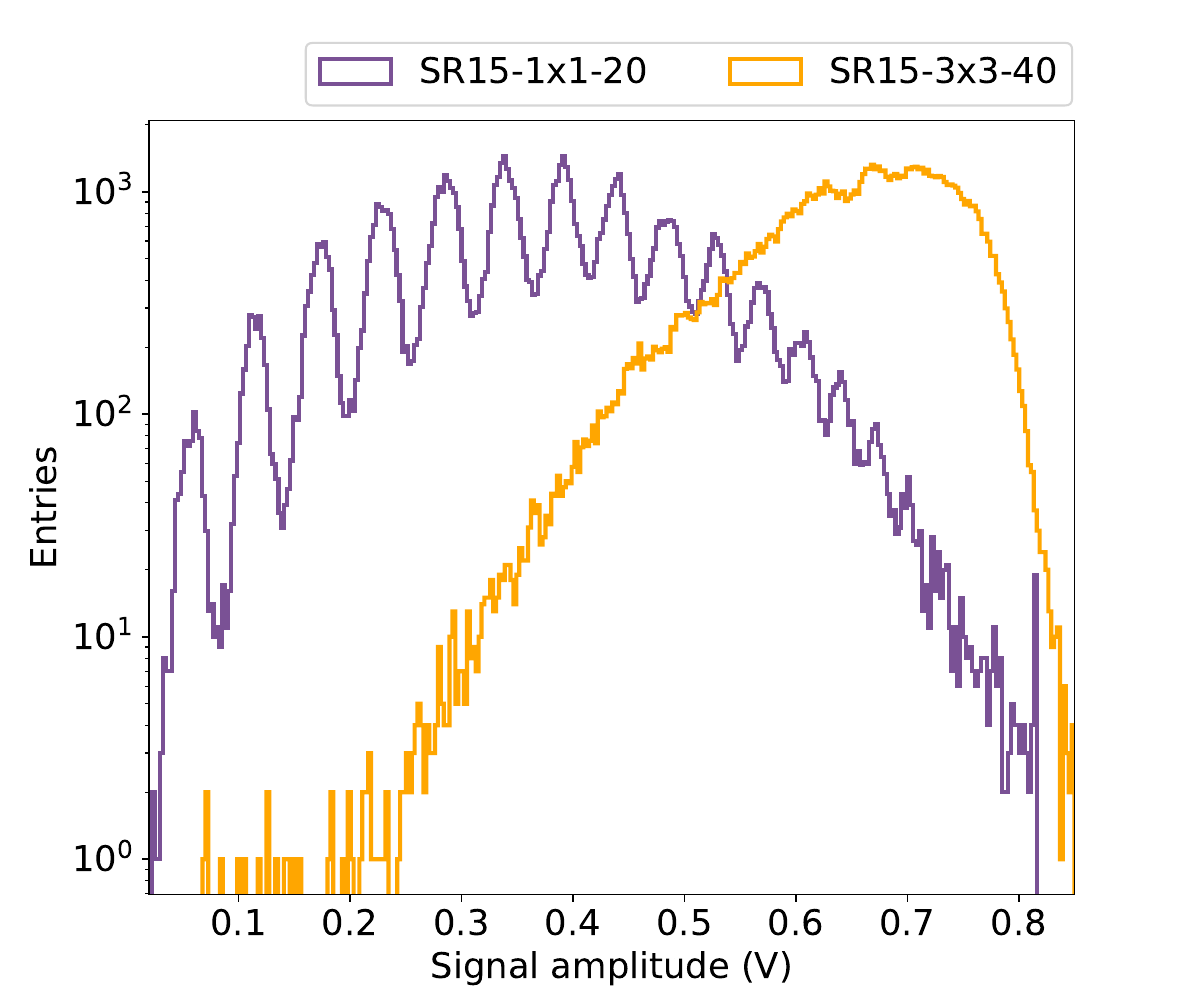}
        \caption{Distribution of the raw signal amplitudes of SR15-1x1-20 with respect to SR15-3x3-40, associated to $10^5$ events at $V_{\text{OV}}~\simeq~2$~V.} 
          \label{fig:LA_signal}
\end{figure}

In Fig. \ref{fig:LA_signal} the distribution of the SR15-3x3-40 signal amplitudes at $V_{\text{OV}}~\simeq~2$~V is reported
; the SR15-1x1-20 signal amplitudes  \cite{SiPM3} are also reported for comparison.
Although in the SR15-3x3-40 the SPADs have larger pitch, i.e. potentially more photons can impinge on a single cell, the average signal is bigger than SR15-1x1-20. This is due to the larger SiPM area and FF (Fill Factor) that allow to cover all the Cherenkov cone subtended area.  In the same figure, it can be noticed that in SR15-1x1-20  the contributions of the different number of fired SPADs is very clear, while for SR1-3x3-40 the larger baseline fluctuations together with the amplifier saturation spoil the separation.

\begin{figure}[h!]
        \centering%
        \includegraphics [width=1\textwidth]{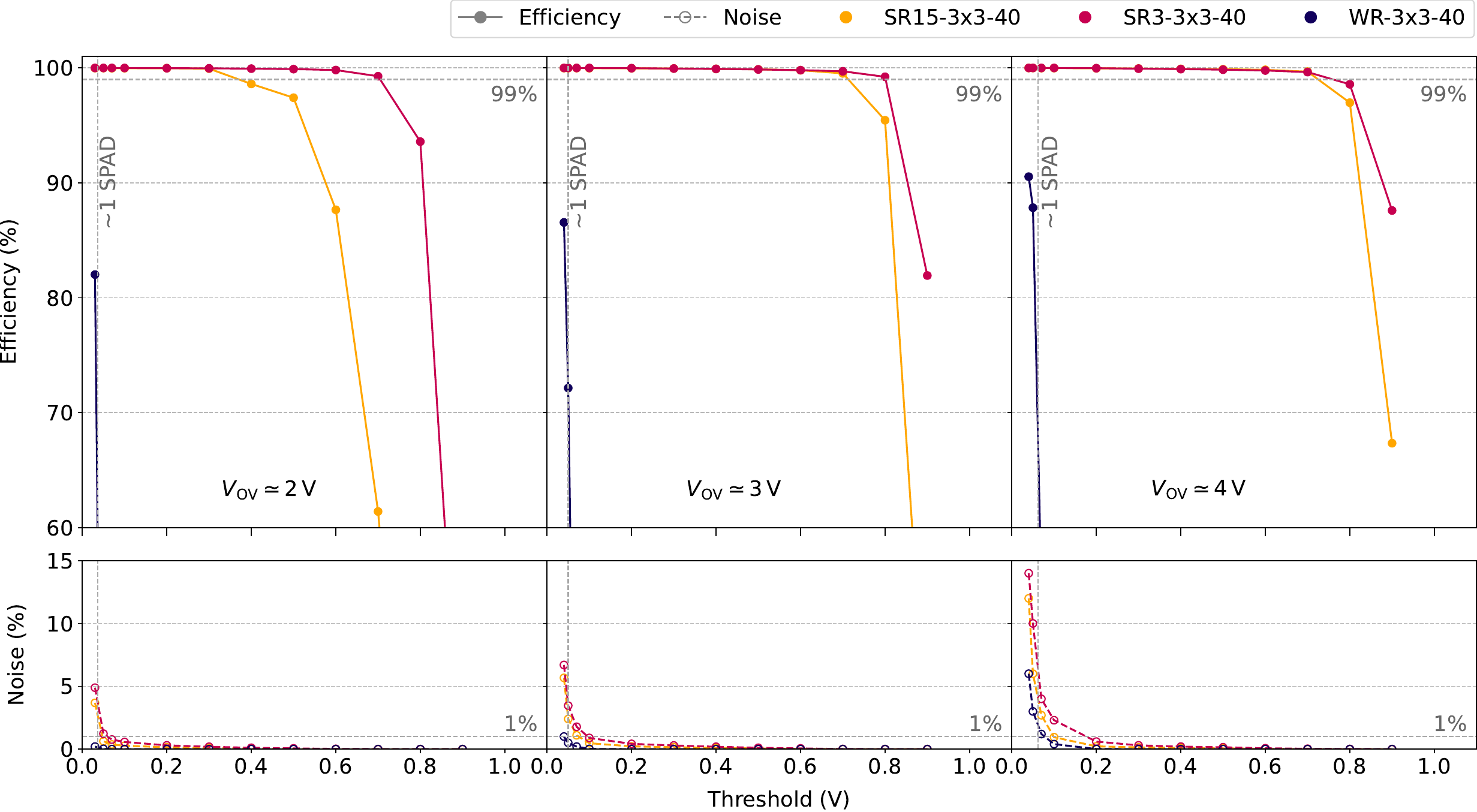}
        \caption{Efficiency versus signal threshold for different 3x3-40 SiPMs, with and without protection,  at different $V_\text{OV}$. The equivalent measurements of noise in a region prior to the signal is  reported in the lower panels of the figure. One SPAD signal, reported with vertical dashed lines, is around $36$ mV, $50$ mV and $62$ mV at 2, 3 and 4 $V_{\text{OV}}$ respectively.} 
          \label{fig:LA_efficiency}
\end{figure}

In the top and bottom panel of Fig. \ref{fig:LA_efficiency} the efficiency and noise of 3x3-40 SiPMs as a function of the applied threshold are reported for different $V_{\text{OV}}$ and protection layer thicknesses. 
The signal amplitude of one SPAD is $\sim$36, 50 and 62 mV for 2, 3 and 4 V OV respectively.\\
The results indicate that SiPMs with a protection layer maintain efficiencies above 99$\%$ while keeping noise contamination below $\sim$1\% over a broad thresholds range for all $V_{\text{OV}}$. The operational ranges for each SiPM and voltage are reported in Table \ref{tab:opranges}.
\begin{table}
\centering
\caption{Operational threshold ranges for each $V_\text{OV}$ and SiPM allowing noise $<1\%$ and efficiency $>99\%$.}
\label{tab:opranges}
\begin{tabular}{ccc}
\toprule
\textbf{$V_\text{OV}$} & \textbf{SR15-3x3-40} & \textbf{SR3-3x3-40} \\
\midrule
2 V & 0.05 -- 0.4 V & 0.05 -- 0.7 V \\
3 V & 0.07 -- 0.7 V & 0.1 -- 0.8 V \\
4 V & 0.1 -- 0.7 V & 0.2 -- 0.8 V \\
\bottomrule
\end{tabular}
\end{table}
At $V_{\text{OV}}\simeq 2$ V, a threshold between 1 to 2 SPADs is sufficient to reach this operating point, while at higher overvoltages the same performance is achieved at slightly larger thresholds
, between 1 to 3 SPADs for both sensors.
The operational range extends then up to several SPADs
, between 11 to 19 SPADs, for both sensors at various overvoltages.
These results highlight the strong capability of both SiPMs to efficiently directly detect charged particles with very low noise contamination in a large range of operation.\\
At lower overvoltage, the differences between the two devices  at increasing threshold values depend on the different thicknesses of their protection layers; with increasing $V_{\text{OV}}$ these differences are attenuated due to signal saturation (related to the amplifier), leading the efficiency curves to converge and drop sharply around thresholds of 0.85~V.\\
Notice also that, as suggested by previous publications, the sensor without protection shows very low efficiency in detecting charged particles: indeed, the maximum efficiency of $\sim 91\%$ is obtained at $V_{\text{OV}} \simeq 4$ V and at very low threshold with a noise contamination of $\sim 5\%$, far from the mentioned operating point.

\begin{figure}[h!]
        \centering%
        \includegraphics [width=1\textwidth]{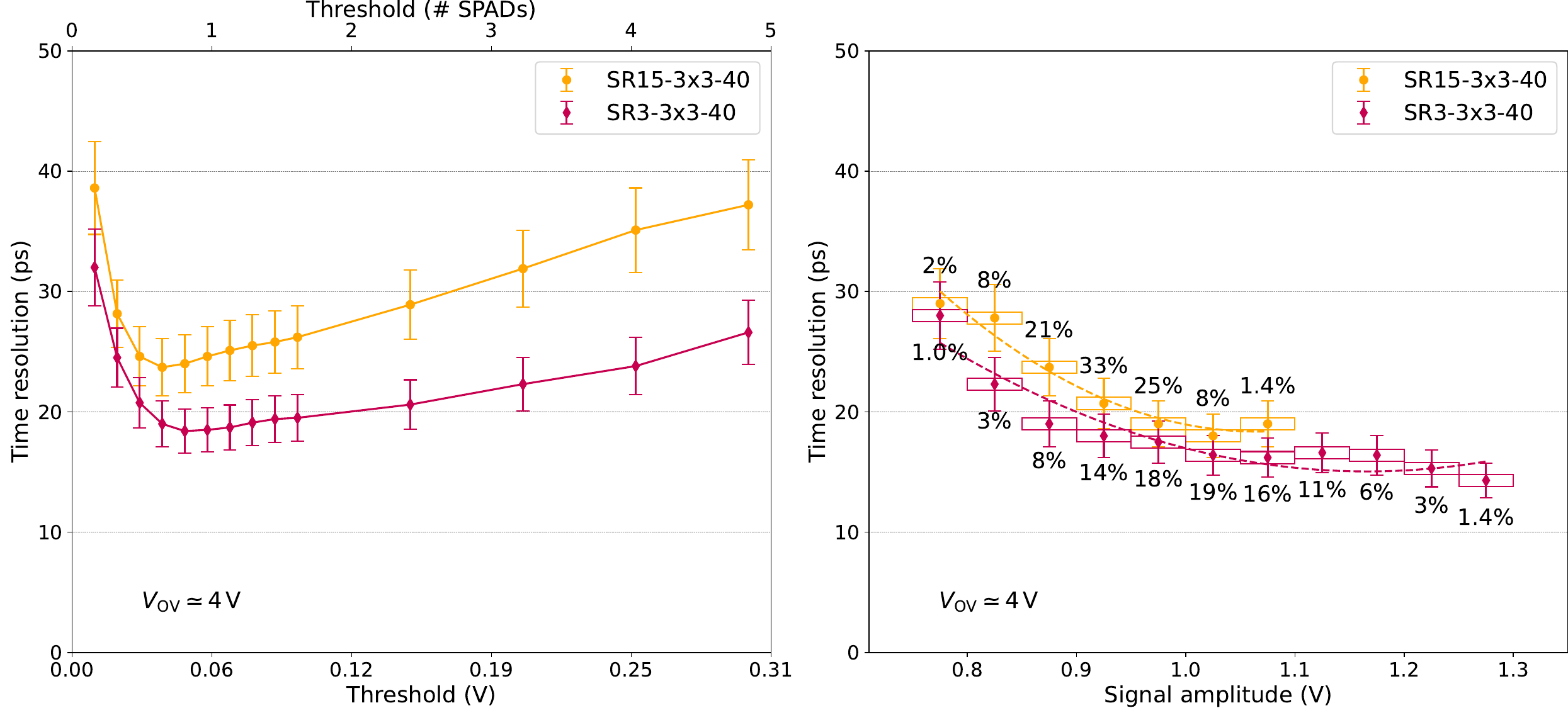}
        \caption{(Left) Time resolution versus timing threshold at $V_\text{OV} \simeq 4$ V for the SiPMs described in Section \ref{sec2}, after the event selection described in the text (on the top x-axis is reported also the number of SPADs corresponding to a given threshold). (Right) Time resolution versus amplitude intervals, considering only the best timing threshold from the left figure. For each point, the corresponding approximate event fraction is shown. The quoted errors of about $10\%$ include the statistics and an evaluation of the systematic uncertainties also considering the reference LGAD contribution.
        } 
          \label{fig:Timing}
\end{figure}

For the timing analysis, the minimum threshold for the signal amplitude was set in order to have an efficiency larger than  $99\%$\footnote{Larger than 0.1 V for SR15-3x3-40 and 0.2 V for SR3-3x3-40 at $V_\text{OV} \simeq 4$ V.}; this threshold corresponds to a signal larger than 2–3 SPADs.
The resulting time resolution as a function of the applied timing threshold at $V_\text{OV} \simeq 4$ V is shown in the left panel of Fig. \ref{fig:Timing}.\\
It can be noticed that for such timing studies the optimal threshold value is lower than that required to achieve $>99\%$ efficiency and $<1\%$ noise contamination, as previously demonstrated.
This observation suggests a possible dual-threshold approach to optimize separately efficiency and time resolution. However, in a single threshold scenario, the same figure indicates that, using the above mentioned minimum threshold necessary  to achieve $>99\%$ efficiency and $<1\%$ noise contamination, would result in a degradation of time resolution by a factor of approximately 1.3, still acceptable depending on the specific requirements of the application.\\
The data confirm that 20 ps time resolution can be achieved at $V_\text{OV} \simeq 4$ V for 3x3-40 SiPMs, similar to the results obtained in \cite{SiPM3} for 1x1-20 SiPMs. This is obtained in a quite wide timing threshold range, making the measurement very stable toward threshold variations. The 3 mm protection layer is slightly better than the 1.5 mm although comparable. 


In the right panel of Fig. \ref{fig:Timing} the time resolution is plotted versus the signal amplitude which is proportional to the number of SPADs fired (as already remarked, for these sensors it is not possible to clearly identify the SPAD peaks). The measurement is performed at a timing threshold corresponding to the minimum of the left plot of Fig. \ref{fig:Timing} ($\sim 60\%$ of 1 SPAD). As already reported in \cite{SiPM3}, with increasing amplitude the resolution improves also for the 3x3-40 sensors. The fraction of selected events falling within each amplitude interval is reported in $\%$ above/below the corresponding point.


The results of this section demonstrate what was noticed in previous results \cite{SiPM3}: the passage of a charged particle through the SiPM produces a very large signal with 100\% efficiency also at high thresholds, thus allowing a safe operation margin with respect to the background. Also, the timing performance is in line with the expectations and may be further improved with optimized front-end electronics.

\subsection{Irradiated samples}
\label{subsec:irrad}


In the same test beam we investigated the DCR and the time resolution before and after sensor irradiation with  10$^{10}$~1~MeV~n$_\text{eq}$~cm$^{-2}$. The 
SiPM under test was one with standard 1 mm silicone resin, named SR1-1x1-20, already used in \cite{SiPM3}.

When the SiPMs were new, the DCR was measured to be $\sim5~10^4$ Hz/mm$^2$ at 4 $V_{\text{OV}}$, consistent with the literature \cite{Acerbi}. 
During the first test beam, the sensor accumulated an irradiation dose estimated to be on the order of $2-3~$10$^{9}$~1~MeV~n$_\text{eq}$~cm$^{-2}$. This accumulated dose was determined based on the total time on beam, particle flux, and momentum and was later found to be consistent with the expected increase in the DCR for this dose range \cite{Altamura}. This dose represents the condition \textit{before} the TIFPA irradiation\footnote{Note that the SiPMs studied in Section \ref{subsec:efficiency} accumulated a comparable dose, which may account for any minor performance degradation w.r.t. brand-new sensors.}.\\

\begin{figure}[h!]
        \centering%
        \includegraphics [width=0.7\textwidth]{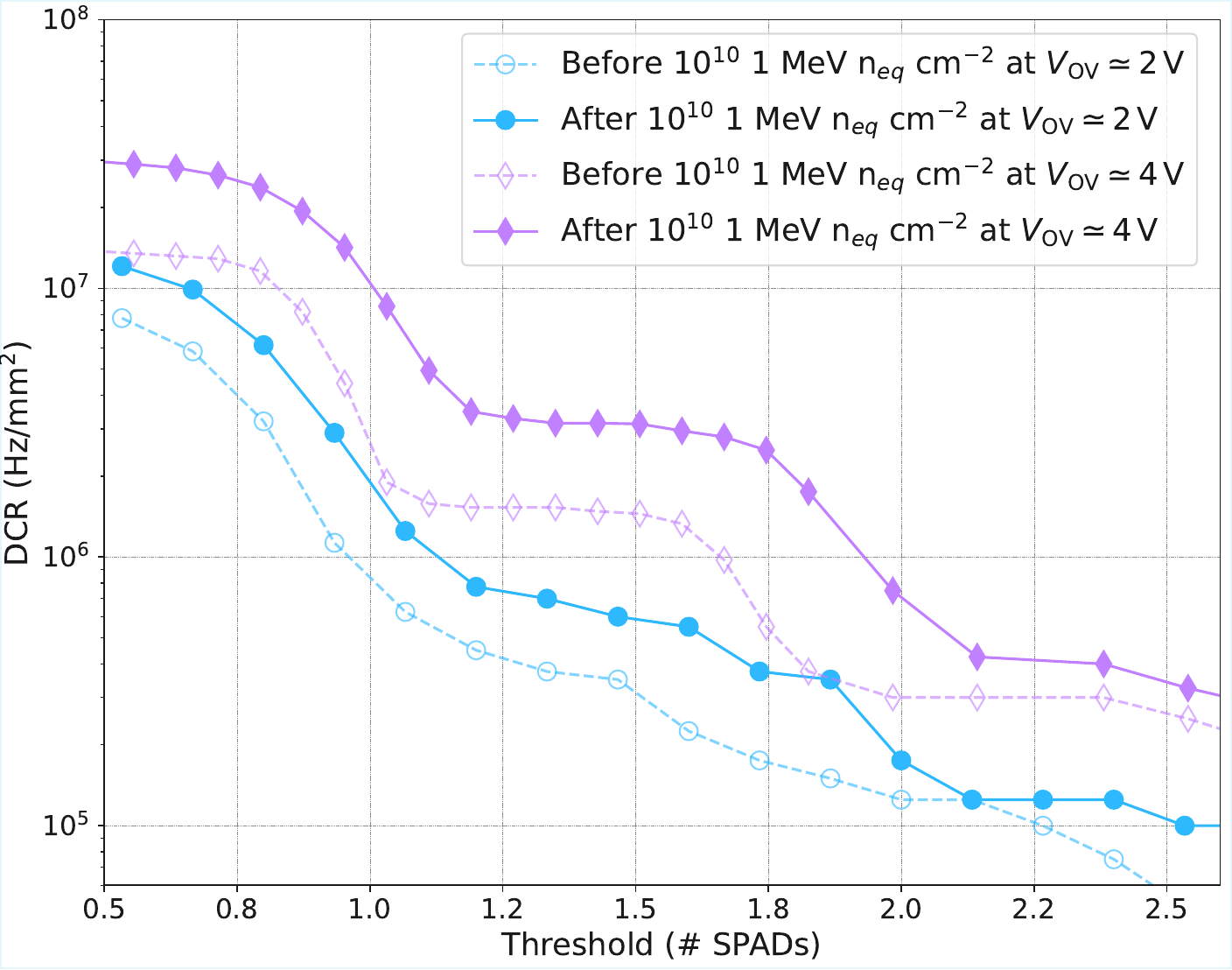}
        \caption{DCR versus threshold for SR1-1x1-20 at $V_{\text{OV}}$ of 2 V and 4 V before and after irradiation. Further details are provided in the text.} 
          \label{fig:RAD_DCR}
\end{figure}

The DCR at a $V_\text{OV}$ of 2~V and 4~V  before and after receiving 10$^{10}$~1~MeV~n$_\text{eq}$~cm$^{-2}$ are reported in Fig. \ref{fig:RAD_DCR} as a function of the applied threshold expressed in terms of associated number of SPADs. 
Fig. \ref{fig:RAD_DCR} demonstrates that the DCR of the irradiated sensor can be reduced by a factor $2-4$~10$^2$ depending on the applied $V_\text{OV}$, still maintaining a very good efficiency for charged particles since the signals are well above 2-3 SPADs, see Fig. \ref{fig:LA_signal} and \cite{SiPM3}. 





\begin{figure}[h!]
        \centering%
        \includegraphics [width=1\textwidth]{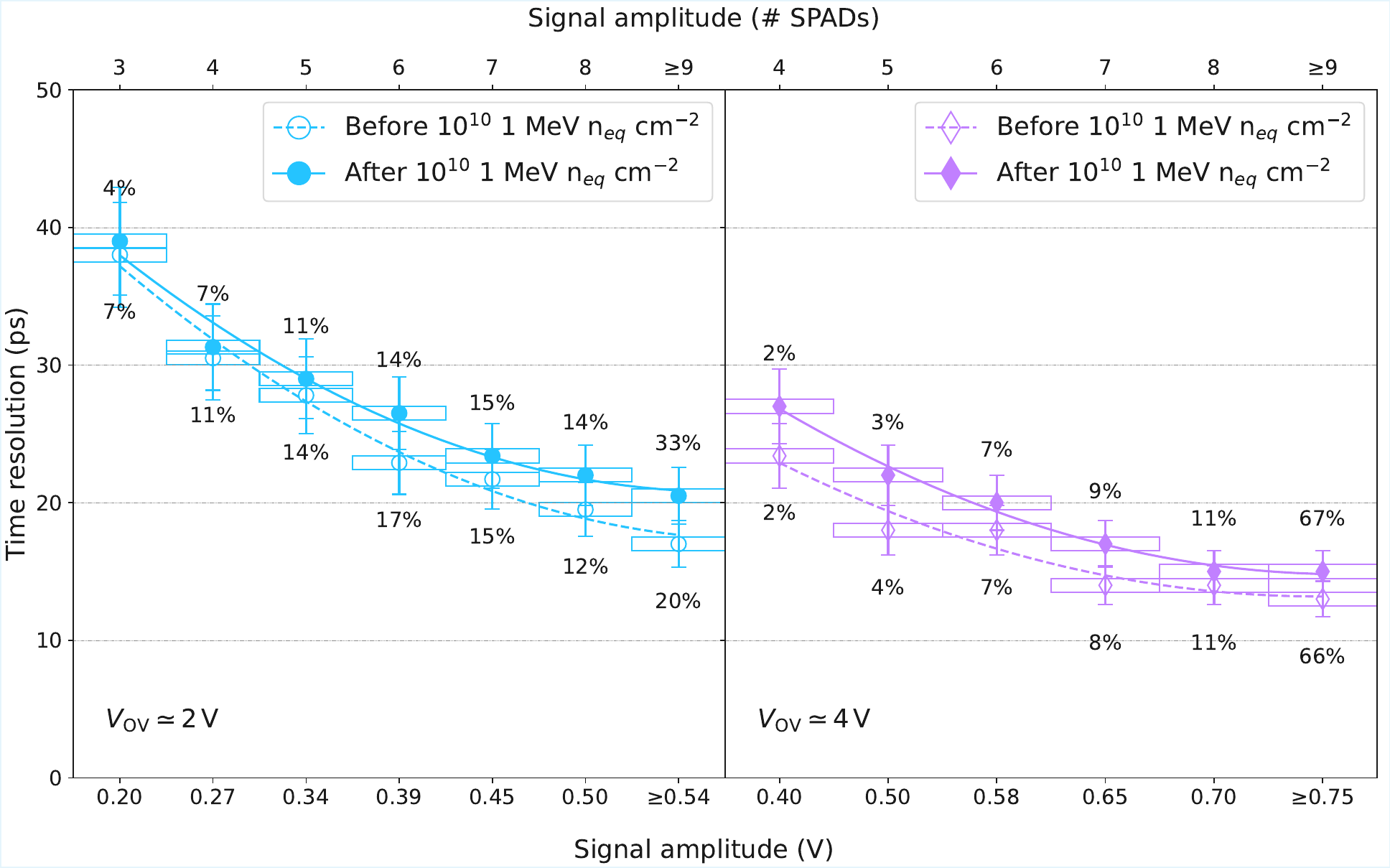}
       \caption{Time resolution of SR1-1x1-20 before and after irradiation versus number of fired SPADs at $V_{\text{OV}}$ of 2 V (left) and 4 V (right). On the bottom x-axis the corresponding mean peak value of every SPAD is reported. For each point the approximated corresponding fraction of events is reported. Errors are evaluated as in Fig. \ref{fig:Timing}. Further details are provided in the text.
       }
          \label{fig:RAD_timing}
\end{figure}

Fig. \ref{fig:RAD_timing} compares the time resolution versus the number of fired SPADs before and after irradiation at two different $V_{\text{OV}}$.  
Non irradiated and irradiated sensors were compared with the same analysis described in Section \ref{subsec:signal}. 
For this analysis the threshold was set to 35$\%$ and 50$\%$ of the single SPAD signal respectively at $V_{\text{OV}}$ of 2 V and 4 V, a value much lower than the signal amplitude (as done for 3x3 SiPMs, reported in Fig. \ref{fig:Timing}).

The results of this section indicate that, after irradiation, the DCR can be reduced by applying a threshold well above a single SPAD, without affecting the detection efficiency, as inferred from the results presented in Section \ref{subsec:efficiency}.
Moreover, timing performances show no significant variation after irradiation, both at $V_{\text{OV}}$ of 2~V and 4~V.



By comparing Fig. 3 (right) and Fig. 5 (right), it can be observed that the achieved time resolutions are compatible. However, when compared at the same signal amplitude (i.e. same number of fired SPADs), the SiPM technology 1×1-20 appears to perform better than the 3×3-40 device, reaching lower time resolution. Although these two SiPMs differ in several design aspects, this behavior can most likely be attributed to the smaller active area of the 1×1 devices, which results in a lower capacitance, in line with the literature. On the other hand, the larger area of the 3×3 devices enables the collection of a higher number of SPADs, thus reaching higher amplitudes. It is therefore noteworthy that, despite these differences, both technologies eventually converge to comparable time resolutions.

\section{Conclusions}\label{conclusion}

In this paper, results on the direct detection of charged particles with SiPMs have been presented.

Efficiency and timing measurements were performed on sensors with a $3\times3$~mm$^2$ active area and a $40$~\textmu m SPAD pitch. 
An efficiency near $100\%$ has been observed up to high thresholds, well above the single SPAD signal.
A time resolution of less than $20$~ps has also been measured, improving with the signal amplitude (i.e. with the number of fired SPADs).

Thanks to the large signal amplitudes detected on the SiPM at the passage of a charged particle, time resolution is not degraded by irradiation, within uncertainties, up to 10$^{10}$~1~MeV~n$_\text{eq}$~cm$^{-2}$. 
Furthermore, the increase in DCR induced by irradiation can be mitigated by applying a sufficiently high threshold, with a negligible impact on the detection efficiency.



\section*{Declarations}
This project has received funding from INFN, FBK and the European Unions Horizon Europe research and innovation programme under grant agreement No 101057511.
The authors received research support from institutes as specified in the author list below the title. \\

\section*{Data availability}
The datasets generated and/or analysed during the current study are available from the corresponding authors on reasonable request.

\bibliography{sn-bibliography}


\end{document}